\documentclass[aps,prd,twocolumn,superscriptaddress,nofootinbib]{revtex4}

\newcommand{\ba}{\begin{array}}

\newcommand{\ea}{\end{array}}

\newcommand{\beqa}{\begin{eqnarray}}

\newcommand{\eeqa}{\end{eqnarray}}

\usepackage[dvips]{color}

\usepackage{fancyhdr}

\usepackage{enumerate}

\usepackage{amsmath,amsthm,amssymb}

\usepackage{slashed}

\usepackage{array}

\usepackage{latexsym}

\newtheorem{theorem}{Theorem}

\usepackage{mathrsfs,bbm}

\begin{document}

\title{Stability of the Scalar Potential and Symmetry Breaking  in the Economical 3-3-1 Model}

\author{Yithsbey Giraldo} 

\affiliation{Instituto de F\'\i sica, Universidad de Antioquia,
A.A. 1226, Medell\'\i n, Colombia.}

\affiliation{Departamento de F\'\i sica, Universidad de Nari\~no, A.A. 1175, Pasto, Colombia.}

\author{William A. Ponce}

\affiliation{Instituto de  F\'\i sica, Universidad de Antioquia,
A.A. 1226, Medell\'\i n, Colombia.}

\author{Luis A. S\'anchez}

\affiliation{Escuela de F\'\i sica, Universidad Nacional de Colombia,
A.A. 3840, Medell\'\i n, Colombia.}

\date{\today}

%\maketitle

\begin{abstract}

A detailed study of the criteria for stability of the scalar potential and the proper electroweak symmetry breaking pattern in the economical 3-3-1 model, is presented. For the analysis we use, and improve, a method previously developed to study the scalar potential in the two-Higgs-doublet extension of the standard model. A new theorem related to the stability of the potential is stated. As a consequence of this study, the consistency of the economical 3-3-1 model emerges.

\end{abstract}


\maketitle

%%%%%%%%%%%%%%%%%%%%%%%%%%%%%%%%%%%%%%%%%%%%%%%%%%%%%%%%%%%%%%%%%%%%%%%%%%%%%%%%%%%%%%%%%%%%%
\section{Introduction}
Extensions of the standard model (SM) based on the local gauge group
$SU(3)_c\otimes SU(3)_L\otimes U(1)_X$~\cite{pf,vl,ozer,sher,pfs} (called hereafter 3-3-1 for short) contain, in general, a scalar sector quite complicated to be analyzed in detail. For this type of models, three Higgs triplets, and in some cases one additional Higgs sextet are used, in order to break the symmetry and provide at the same time with masses to the fermion fields of each model~\cite{rob}.

Among the 3-3-1 models with the simplest scalar sector are the ones proposed for the first time in Ref.~\cite{b3} and further analyzed in Refs.~\cite{b4} (they make use of only two scalar Higgs field triplets). This class of models include eight different three-family models where the Higgs scalar fields, the gauge-boson sector and the fermion field representations are restricted to particles without exotic electric charges~\cite{pfs,b3}. Because of their minimal content of Higgs scalar fields they are named in the literature ``economical 3-3-1 models".

A simple extension of the SM consists of adding to the model a second Higgs scalar doublet~\cite{dona}, defining in this way the so-called two-Higgs-doublet model (THDM). The different ways how the two Higgs scalar doublets couple to the fermion sector define the several versions of this extension\cite{dona,b5}. Many gauge group extensions of the SM have the THDM as an effective low energy theory (in this regard see the papers in \cite{b5} and references therein).
In these extensions one of the first steps in the symmetry breaking chain leads to the $SU(3)_c\otimes SU(2)_L\otimes U(1)_Y$ gauge theory with two Higss doublets in one of its several versions.

A novel method for a detailed analysis of the scalar potential in the most general THDM  was presented in Refs.~\cite{b1} where by using powerful algebraic techniques, the authors studied in detail the stationary points of the scalar potential. This allowed them to give, in a very concise way, clear criteria for the stability of the scalar potential and for the correct electroweak symmetry breaking pattern. In the present work we use this approach to analyse the scalar sector of the economical 3-3-1 model. No relevant new additional conditions are necessary to be imposed in order to implement the method in this last case.

One important advantage of the economical 3-3-1 model, compared with the THDM, concerns the Higgs potential. The 14 parameters required to describe the most general potential for the second case, should be compared with the six parameters required in the economical 3-3-1 model. For the THDM this is associated to the fact that the two Higgs doublets have the same $U(1)$ hypercharge~\cite{dona,b5}. In the economical 3-3-1 model, by contrast, the two scalar triplets have different $U(1)_X$ hypercharges so that the most general Higgs potential shows itself in a very simple form.

In this work we deduce constraints on the parameters of the economical 3-3-1 scalar potential coming from the stability and from the electroweak symmetry breaking conditions. The stability of an scalar potential at the classical level, which is fulfilled when it is bounded from below, is a necessary condition in order to have a sound theory. The global minimum of the potential is found by determining its stationary points. Some of our results agree with those already presented in Refs.~\cite{b3,b4}. Our study extends thus the method proposed in Refs.~\cite{b1,b2} to the economical  3-3-1 model, where the results are very concise and should, in principle, be used as a guide in order to extend the method to other situations.

This paper is organized as follows: in Sect.~\ref{sec:sec2} we briefly review the mathematical formalism in order to make this work self-contained; in Sect.~\ref{sec:sec3} we apply the method to the scalar sector of the economical 3-3-1 model, which  is followed in
Sect.~\ref{sec:sec5} by the introduction of new parameterizations. In Sect.~\ref{sec:sec6} we derive expressions for the masses of the scalar fields, and our conclusions are presented in Sect.~\ref{sec:sec7}. In Appendix~\ref{sec:secA1} a new theorem that facilitates the stability criteria is proved. In Appendix~\ref{sec:secA2} two exceptional solutions for the global minimum of the potential are analyzed. Finally, in Appendix~\ref{apendice3}, it is verified that if only one scalar triplet acquires a nonzero Vacuum Expectation Value (VEV), the economical 3-3-1 model is inconsistent.

%%%%%%%%%%%%%%%%%%%%%%%%%%%%%%%%%%%%%%%%%%%%%%%%%%%%%%%%%%%%%%%%%%%%%%%%%%%%%%%%%%%%%%%%%%%%%

\section{\label{sec:sec2}A review of the method}

In this section, and following Refs. \cite{b1} and \cite{b2}, we review a new algebraic approach used to determine the global minimum of the Higgs scalar potential, its stability, and the spontaneous symmetry breaking from $SU(2)_L\otimes U(1)_Y$ down to $U(1)_{em}$, in the extension of the SM known as the THDM, where $\varphi_1$ and $\varphi_2$ stand for two Higgs scalar field doublets with identical quantum numbers

Stability and the stationary points of the potential can be analyzed in terms of four real constants given by
\begin{equation}
\label{8}
 K_0=\sum_{i=1,2}\varphi_i^\dag\varphi_i,\quad K_a=\sum_{i,j=1,2}(\varphi_i^\dag\varphi_j)\sigma^a_{ij},\quad (a=1,2,3).
\end{equation}
where $\sigma^a(a=1,2,3)$ are the Pauli spin matrices. The four vector $(K_0,\boldsymbol K)$ must lie on or inside the forward \textit{light cone}, that is
\begin{equation}
\label{19}
 K_0\geq 0,\quad K_0^2-\boldsymbol K^2\geq 0.
\end{equation}
Then the positive and hermitian $2\times 2$ matrix
\begin{equation}
\label{6aa}
\underline K=
\begin{pmatrix}
             \varphi_1^\dag\varphi_1 & \varphi_2^\dag\varphi_1\\
\varphi_1^\dag\varphi_2 & \varphi_2^\dag\varphi_2
              \end{pmatrix}
\end{equation}
may be written as
\begin{equation}\label{6ab}
\underline K_{ij}=\frac{1}{2}(K_0\delta_{ij}+K_a\sigma^a_{ij}).
\end{equation}
Inverting Eq.~(\ref{8}) it is obtained
\begin{equation}
\begin{split}
 \varphi_1^\dag\varphi_1=(K_0+K_3)/2,\quad  \varphi_1^\dag\varphi_2=(K_1+iK_2)/2,\\
 \varphi_2^\dag\varphi_2=(K_0-K_3)/2,\quad  \varphi_2^\dag\varphi_1=(K_1-iK_2)/2\:.
\end{split}
\end{equation}
The most general $SU(2)_L\otimes U(1)_Y$ invariant Higgs scalar potential can thus be expressed as
\begin{subequations}
\label{10aa}
\begin{align}
 V(\varphi_1,\varphi_2)&=V_2+V_4,\\
V_2&=\xi_0K_0+\xi_aK_a,\\
V_4&=\eta_{00}K_0^2+2K_0\eta_aK_a+K_a\eta_{ab}K_b,
\end{align}
\end{subequations}
where the 14 independent parameters $\xi_0,\;\xi_a,\; \eta_{00},\; \eta_{a}$ and $\eta_{ab}=\eta_{ba}$ are real. Subsequently, it is defined $\boldsymbol K=(K_a),\; \boldsymbol{\xi}=(\xi_a),\; \boldsymbol\eta=(\eta_a)$ and $E=(\eta_{ab})$.

%%%%%%%%%%%%%%%%%%%%%%%%%%%%%%%%%%%%%%%%%%%%%%%%%%%%%%%%%%%%%%%%%%%%%%%%%%%%%%%%%%%%%%%%%%%

%%%%%%%%%%%%%%%%%%%%%%%%%%%%%%%%%%%%%%%%%%%%%%%%%%%%%%%%%%%%%%%%%%%%%%%%%%%%%%%%%%%%%%%%%%%

\subsection{\label{sec:sec21}Stability}
From (\ref{10aa}), for~\mbox{$K_0 > 0$} and defining $\boldsymbol{k} = \boldsymbol{K} / K_0$, it is obtained
\begin{align}
\label{eq-vk}
V_2 &= K_0\, J_2(\boldsymbol{k}),&
J_2(\boldsymbol{k}) &:= \xi_0 + \boldsymbol{\xi}^\mathrm{T} \boldsymbol{k},\\
\label{eq-vk4}
V_4 &= K_0^2\, J_4(\boldsymbol{k}),&
J_4(\boldsymbol{k}) &:= \eta_{00}
  + 2 \boldsymbol{\eta}^\mathrm{T} \boldsymbol{k} + \boldsymbol{k}^\mathrm{T} E \boldsymbol{k},
\end{align}
where the functions $J_2(\boldsymbol{k})$ and $J_4(\boldsymbol{k})$
on the domain \mbox{$|\boldsymbol{k}| \leq 1$} have been introduced.
For the potential to be stable, it must be bounded from below.
The stability is determined by the behavior of $V$ in the limit
\mbox{$K_0 \rightarrow \infty$}, and hence by the signs of
\mbox{$J_4(\boldsymbol{k})$} and \mbox{$J_2(\boldsymbol{k})$} in~(\ref{eq-vk}) and~(\ref{eq-vk4}).
In this analysis only the \emph{strong} criterion for stability is considered, that is, the stability is determined solely by the $V$ quartic terms
\begin{equation}
\label{eq-jinq}
J_4(\boldsymbol{k}) > 0 \quad \text{for all }\left\lvert {\boldsymbol{k}}\right\rvert \leq 1.
\end{equation}
To assure that $J_4(\boldsymbol{k})$ is always positive, it is sufficient to
consider its value for all its stationary points
on the domain \mbox{$\left\lvert {\boldsymbol{k}}\right\rvert < 1$}, and for all the stationary points on the
boundary~\mbox{$|\boldsymbol{k}| = 1$}.
This leads to bounds on $\eta_{00}$, $\eta_{a}$ and $\eta_{ab}$, which
parameterize the quartic term~$V_4$ of the potential.

The regular solutions for the two cases \mbox{$\left\lvert {\boldsymbol{k}}\right\rvert < 1$} and
\mbox{$\left\lvert {\boldsymbol{k}}\right\rvert = 1$} lead to
\begin{align}
\label{eq-flam}
f(u) & =  u + \eta_{00} - \boldsymbol{\eta}^\mathrm{T} (E - u)^{-1} \boldsymbol{\eta},\\
\label{eq-flampr}
f'(u) & =  1 - \boldsymbol{\eta}^\mathrm{T} (E - u)^{-2} \boldsymbol{\eta},
\end{align}
so that for all ``regular'' stationary points~$\boldsymbol{k}$
of~$J_4(\boldsymbol{k})$ both
\begin{align}
\label{11d}
f(u) &= \left. J_4(\boldsymbol{k}) \right|_{\substack{stat}},\quad \mbox{and}\\
f'(u) &= 1 - \boldsymbol{k}^2
\end{align}
hold, where $u=0$ must be set for the solution with $\left\lvert {\boldsymbol{k}}\right\rvert<1$.
There are stationary points of~$J_4(\boldsymbol{k})$ with $\left\lvert {\boldsymbol{k}}\right\rvert<1$
and $\left\lvert {\boldsymbol{k}}\right\rvert=1$ exactly if $f'(0)>0$ and $f'(u)=0$, respectively,
and the value of $J_4(\boldsymbol{k})$ is then given by $f(u)$.

In a basis where \mbox{$E = {\rm diag}(\mu_1, \mu_2, \mu_3)$} it is obtained
\begin{align}
\label{eq-fdiag}
f(u) &= u + \eta_{00} - \sum_{a = 1}^3 \frac{\eta_a^2}{\mu_a - u},\\
\label{eq-fprd}
f'(u) &= 1 - \sum_{a = 1}^3 \frac{\eta_a^2}{(\mu_a - u)^2}.
\end{align}
The derivative~\mbox{$f'(u)$} has at most six zeros.
Notice that there are no exceptional solutions
if in this basis all three components of~$\boldsymbol{\eta}$ are different from zero.

Consider now the functions $f(u)$ and $f'(u)$ and denote by $I$
\begin{equation}
\label{eq-idef}
I = \{ u_1, \dots, u_n \}
\end{equation}
the set of values $u_j$ for which $f'(u_j)=0$. Add $u_k=0$ to $I$ if $f'(0)>0$.
Consider then the eigenvalues $\mu_a$ ($a=1,2,3$) of $E$.
Add those $\mu_a$ to $I$ where $f(\mu_a)$ is finite and $f'(\mu_a) \geq 0$.
Then $n \leq 10$. The values of the function
$J_4(\boldsymbol{k})$
at its stationary points are given by
\begin{equation}
\left. J_4(\boldsymbol{k}) \right|_{\substack{stat}} = f(u_i)
\end{equation}
with $u_i \in I$. In Appendix A we show that the stationary point in $I$ having the smallest value, will produce the smallest value of $J_4(\boldsymbol k)$ in the domain $|\boldsymbol k|\leq1$. We now state the theorem.
%
%\begin{theopargself}
\begin{theorem}
\label{t1}
The global minimum of the function $J_4(\boldsymbol k)$, in the domain $|\boldsymbol k|\leq1$,  is given and guaranteed by the stationary point of the set $I$ with the smallest value.
\end{theorem}
%\end{theopargself}
%
This result guarantees strong stability if $f(u)>0$, where $u$ is the smallest value of $I$.
The potential is unstable if we have $f(u)<0$. If $f(u)=0$ we have to consider in addition $J_2(\boldsymbol k)$ in order to decide on the stability of the potential.

%%%%%%%%%%%%%%%%%%%%%%%%%%%%%%%%%%%%%%%%%%%%%%%%%%%%%%%%%%%%%%%%%%%%%%%%%%%
\subsection{\label{sec:sec22}Location of stationary points and criteria for electroweak symmetry breaking}
The next step after the stability analysis in the preceding section has been done is to determine the
location of the stationary points of the potential, since among these points
the local and global minima are found. To this end is defined
\begin{equation}\label{tildepar}
\boldsymbol{\tilde{K}} = \begin{pmatrix} K_0\\ \boldsymbol{K} \end{pmatrix},\quad
\boldsymbol{\tilde{\xi}}  = \begin{pmatrix} \xi_0\\ \boldsymbol{\xi} \end{pmatrix},\quad
\tilde{E} = \begin{pmatrix} \eta_{00} & \boldsymbol{\eta}^\mathrm{T}\\
                           \boldsymbol{\eta} & E \end{pmatrix}.
\end{equation}
In this notation the potential~(\ref{10aa}) reads
\begin{equation}
\label{eq-vtil}
V =\boldsymbol{\tilde{K}}^\mathrm{T}\boldsymbol{\tilde{\xi}}  + \boldsymbol{\tilde{K}}^\mathrm{T}
\tilde{E} \boldsymbol{\tilde{K}}
\end{equation}
and is defined on the domain
\begin{equation}
\label{eq-domv}
\boldsymbol{\tilde{K}}^\mathrm{T} \tilde{g} \boldsymbol{\tilde{K}}\geq 0,
\qquad K_0 \ge 0,
\end{equation}
with
\begin{equation}
\tilde{g} = \begin{pmatrix} 1 & \phantom{-}0 \\ 0 & -\mathbbm{1} \end{pmatrix}.
\end{equation}
For the discussion of the stationary points of~$V$, three different cases must be distinguished:
$\boldsymbol{\tilde{K}}=0$,
$K_0 > \left\lvert {\boldsymbol{K}} \right\rvert$, which are the solutions inside the forward light cone,
and
$K_0 = \left\lvert {\boldsymbol{K}} \right\rvert > 0$, which are the solutions on the forward light cone.

The trivial configuration $\boldsymbol{\tilde{K}}=0$
is a stationary point of the potential with~$V=0$,
as a direct consequence of the definitions.
The stationary points of~$V$ in the inner part of the domain,
$K_0>\left\lvert {\boldsymbol{K}} \right\rvert$,
are given by
\begin{equation}
\label{eq-statin}
\tilde{E} \boldsymbol{\tilde{K}}  = - \frac{1}{2} \boldsymbol{\tilde{\xi}},
\quad
\text{with}
\quad
\boldsymbol{\tilde{K}}^\mathrm{T} \tilde{g} \boldsymbol{\tilde{K}}>0
\quad
\text{and}
\quad
K_0>0.
\end{equation}
The stationary points of $V$ on the domain boundary
$K_0 = \vert \boldsymbol{K}\vert > 0$ are stationary points of the function
\begin{equation}
\tilde{F}\big(\boldsymbol{\tilde{K}}, w \big) := V - w \boldsymbol{\tilde{K}}^\mathrm{T}
\tilde{g} \boldsymbol{\tilde{K}},
\end{equation}
where $w$ is a Lagrange multiplier. The relevant stationary points of
\mbox{$\tilde{F}$} are given by
\begin{equation}
\label{eq-stap}
\big(\tilde{E} -w \tilde{g} \big) \boldsymbol{\tilde{K}} =
-\frac{1}{2} \boldsymbol{\tilde{\xi}},
\quad \text{with}\quad \boldsymbol{\tilde{K}}^\mathrm{T} \tilde{g} \boldsymbol{\tilde{K}}=0 \quad \text{and}\quad K_0 >0.
\end{equation}
For \emph{any} stationary point the potential is given by
\begin{equation}
\label{eq-statexpl}
V|_{\substack{stat}} = \frac{1}{2} \boldsymbol{\tilde{K}}^\mathrm{T} \boldsymbol{\tilde{\xi}}= -
\boldsymbol{\tilde{K}}^\mathrm{T} \tilde{E} \boldsymbol{\tilde{K}}.
\end{equation}
Similarly to the stability analysis in Sec.~\ref{sec:sec21}, a unified description for the regular stationary points of~$V$ with
$K_0>0$ for both \mbox{$|\boldsymbol{K}| < K_0$} and \mbox{$|\boldsymbol{K}| = K_0$} can be used by defining the functions
\begin{align}
\label{eq-ftil}
\tilde{f}(w) &= -\frac{1}{4} \boldsymbol{\tilde{\xi}}^{\, \rm T}
 \big( \tilde{E} - w \tilde{g} \big)^{-1} \boldsymbol{\tilde{\xi}},\\
\label{eq-ftilpr}
\tilde{f}'(w) &= - \frac{1}{4} \boldsymbol{\tilde{\xi}}^{\, \rm T}
\big( \tilde{E}-w\tilde{g} \big)^{-1} \tilde{g}
  \big( \tilde{E}- w \tilde{g} \big)^{-1} \boldsymbol{\tilde{\xi}}.
\end{align}
Denoting the first component of \mbox{$\boldsymbol{\tilde{K}}(w)$} as $K_0(w)$
the following theorem holds.
\begin{theorem}
\label{t2}
\begin{samepage}
\label{classes-statpoints}
The stationary points of the potential are given by
\begin{itemize}
\item[(I\,a)]
$\boldsymbol{\tilde{K}} = \boldsymbol{\tilde{K}}(0)$
if $\tilde{f}'(0) < 0$,\  $K_0(0)>0$ and \mbox{$\det \tilde{E} \neq 0$},
\item[(I\,b)]
solutions $\boldsymbol{\tilde{K}}$ of~\eqref{eq-statin}
if \mbox{$\det \tilde{E} = 0$},
\item[(II\,a)]
$\boldsymbol{\tilde{K}}=\boldsymbol{\tilde{K}}(w)$
for $w$ with \mbox{$\det (\tilde{E} - w \tilde{g}) \neq 0$},\
 \mbox{$\tilde{f}'(w) = 0$} and
 $K_0(w)>0$,
\item[(II\,b)]
solutions $\boldsymbol{\tilde{K}}$ of~\eqref{eq-stap}
for $w$ with \mbox{$\det (\tilde{E} - w\tilde{g}) = 0$},
\item[(III)]
$\boldsymbol{\tilde{K}}$ = 0.
\end{itemize}
\end{samepage}
\end{theorem}
In what follows it is assumed that the potential is stable.
For parameters fulfilling \mbox{$\xi_0 \geq |\boldsymbol{\xi}|$},
this immediately implies
\mbox{$J_2(\boldsymbol{k})\ge 0$} and hence, from the strong condition~\eqref{eq-jinq},
$V>0$ for all \mbox{$\boldsymbol{\tilde{K}} \neq 0$}.
Therefore for these parameters the global minimum is at~$\boldsymbol{\tilde{K}}=0$.
This leads to the requirement
\begin{equation}
\label{eq-cond}
\xi_0 < |\boldsymbol{\xi}|.
\end{equation}
Also, it is obtained
\begin{equation}
\label{28aa}
\left. \frac{\partial V}{\partial K_0} \right|_{
  \begin{subarray}{l} \boldsymbol{k}\;\text{fixed},\\
                    K_0 = 0 \end{subarray}
} = \xi_0 +  \boldsymbol{\xi}^\mathrm{T} \boldsymbol{k}
  < 0
\end{equation}
for some $\boldsymbol{k}$, i.e.\ the global minimum of $V$ lies
at~\mbox{$\boldsymbol{\tilde{K}} \neq 0$} with
\begin{equation}
\label{29b}
 V|_{\substack{min}} <0.
\end{equation}
Firstly, consider $p_0 = \left\lvert {\boldsymbol{p}} \right\rvert$.
From~\eqref{eq-vtil} and \eqref{eq-stap} it follows that
\begin{equation}
\label{55a}
\left. \frac{\partial V}{\partial K_0} \right|_{
  \begin{subarray}{l} \boldsymbol{K}\;\text{fixed},\\
                      \boldsymbol{\tilde{K}}=\boldsymbol{\tilde{p}} \end{subarray}
}
 = \xi_0 + 2 (\tilde{E}\,\boldsymbol{\tilde{p}})_0
 = 2 w_p\, p_0.
\end{equation}
If $w_p<0$, there are points $\boldsymbol{\tilde{K}}$ with $K_0>p_0$,
$\boldsymbol{K}=\boldsymbol{p}$ and lower potential in the neighborhood of
$\boldsymbol{\tilde{p}}$, which therefore cannot be a minimum.
The conclusion is that in a theory with the required electroweak symmetry breaking~(EWSB) the global minimum must have
a Lagrange multiplier such that $w_0 \geq 0$, and for the THDM, the global minimum lies on the stationary points of the classes $(IIa)$ and $(IIb)$ of theorem 2, with the largest Lagrange multiplier~\cite{b1} (contrary to what happens in the analysis that follows for the economical 3-3-1 model, where the global minimum must fall on the stationary points in classes $(Ia)$ and $(Ib)$).

%%%%%%%%%%%%%%%%%%%%%%%%%%%%%%%%%%%%%%%%%%%%%%%%%%%%%%%%%%%%%%%%%%%%%%%%%%%%%%%%%%%%%%%

\section{\label{sec:sec3}The economical 3-3-1 model}
As mentioned before, there exist a total of eight different economical 3-3-1 models without exotic electric charges, each one with a different fermion structure but with the same gauge-boson sector and the same minimal scalar content (two Higgs triplets)~\cite{b3}. The particular economical 3-3-1 model most extensively studied in the literature has the following anomaly free fermion representations:
\begin{eqnarray*}
\psi_L^{a}&=&(l^{-a},\nu^a,N^{0a})^T_L\sim (1,3^*,-1/3),\\
l^{+a}_L &\sim& (1,1,1),\\
Q_L^{i}&=&(u^i,d^i,D^i)^T_L\sim (3,3,0),\\
Q_L^{1}&=&(d^1,u^1,U)^T_L\sim (3,3^*,1/3),\\
u^{ca}_L&\sim&(3^*,1,-2/3),\;\; d^{ca}_L\sim (3^*,1,1/3),\\
U^c_L&\sim&(3^*,1,-2/3),\;\; D^{ci}_L\sim (3^*,1,1/3),
\end{eqnarray*}
where the numbers inside the parentheses stand for $[SU(3)_c$, $SU(3)_L,U(1)_Y]$ representations, $a=1,2,3$ is a family index and $i=1,2$ is related to two of the three families. $D^i$ and $U$ are three exotic quarks with electric charges
$-1/3, \; -1/3$ and $2/3$, respectively.

%%%%%%%%%%%%%%%%%%%%%%%%%%%%%%%%%%%%%%%%%%%%%%%%%%%%%%%%%%%%%%%%%%%%%%%%%%%%%%%%%%%%%

\subsection{\label{sec:sec31}The scalar sector}
If we pretend to use the simplest $SU(3)_L$ representations in order to
break the symmetry, at least two complex scalar triplets, equivalent to
twelve real scalar fields, are required. The two Higgs scalars (together with their complex conjugates) that may develop nonzero VEV, are
\begin{equation}
\label{60a}
 \phi_1(1,3^*,-\frac{1}{3})=\begin{pmatrix}
                             \phi_1^-\\
\phi_1^{\prime 0}\\
\phi_1^0
                            \end{pmatrix},\quad
\phi_2(1,3^*,\frac{2}{3})=\begin{pmatrix}
                             \phi_2^0\\
\phi_2^{+}\\
\phi_2^{\prime +}
                            \end{pmatrix}.
\end{equation}
Note that, unlike the THDM, these two scalar fields have different $X$ hypercharge. For this reason, a change of basis of the Higgs fields in this model does not have any meaning.

The most general, renormalizable and 3-3-1 invariant scalar potential can thus be written as
\begin{equation}
\begin{split}
\label{2aaa}
V(\phi_1,\phi_2)&=\mu_1^2\phi_1^\dag\phi_1+\mu_2^2\phi_2^\dag\phi_2+\lambda_1(\phi_1^\dag\phi_1)^2+\lambda_2(\phi_2^\dag\phi_2)^2\\
&+\lambda_3(\phi_1^\dag\phi_1)(\phi_2^\dag\phi_2)+\lambda_4(\phi_1^\dag\phi_2)(\phi_2^\dag\phi_1).
\end{split}
\end{equation}
The  simplicity of this potential can be appreciated by noticing first the natural absence of a trilinear scalar coupling and by counting its number of free parameters: only  six.

%%%%%%%%%%%%%%%%%%%%%%%%%%%%%%%%%%%%%%%%%%%%%%%%%%%%%%%%%%%%%%%%%%%%%%%%%%%%%%%%%%%%%%%%%%%%

\subsection{\label{sec:sec32}The orbital variables}
Following the method presented in the previous section, the potential (\ref{2aaa}) can be expressed in terms of the orbital variables $K_0,\; K_1,\; K_2$ and $K_3$ which, for our case, are associated to the real parameters
\begin{align}
\label{62aa}
 \xi_0&=\frac{1}{2}(\mu_1^2+\mu_2^2),\quad \boldsymbol \xi=\begin{pmatrix}
 0\\0\\\frac{1}{2}(\mu_1^2-\mu_2^2)
 \end{pmatrix}, \\
\eta_{00}&=\frac{1}{4}(\lambda_1+\lambda_2+\lambda_3),\\
\label{63aa}
\boldsymbol{\eta}&=\begin{pmatrix}
            0\\
0\\
\frac{1}{4}(\lambda_1-\lambda_2)
\end{pmatrix}
,\: E=\begin{pmatrix}
\frac{\lambda_4}{4}&0&0\\
0&\frac{\lambda_4}{4}&0\\
0&0&\frac{1}{4}(\lambda_1+\lambda_2-\lambda_3)
           \end{pmatrix}.
\end{align}

%%%%%%%%%%%%%%%%%%%%%%%%%%%%%%%%%%%%%%%%%%%%%%%%%%%%%%%%%%%%%%%%%%%%%%%%%%
\subsection{\label{sec:sec321}Stability}
Note that $E$ is a diagonal matrix. Then, we can calculate the functions $f(u)$ and $f'(u)$  directly from Eqs.~(\ref{eq-fdiag}) and (\ref{eq-fprd}). We obtain
\begin{align}
\label{6}
f(u)&=u+\frac{1}{4}(\lambda_1+\lambda_2+\lambda_3)-\frac{(\lambda_1-\lambda_2)^2}{4(\lambda_1+\lambda_2-\lambda_3)-16u},\\
\label{5}
 f^\prime(u)&=1-\frac{(\lambda_1-\lambda_2)^2}{(\lambda_1+\lambda_2-\lambda_3-4u)^2}.
\end{align}
For $\lambda_1\neq\lambda_2$, the solutions of $f^\prime(u)=0$, which determine the stationary points of $J_4(\boldsymbol{k})$ on the boundary $|\boldsymbol{k}|=1$, lead to the Lagrange multipliers
\begin{equation}
 u_1=\frac{1}{4}(2\lambda_1-\lambda_3),\quad u_2=\frac{1}{4}(2\lambda_2-\lambda_3).
\end{equation}
We must add the values
\begin{equation}
 u_3=0,\quad u_4=\frac{\lambda_4}{4},
\end{equation}
which correspond to the stationary point inside the sphere $\left(|\boldsymbol{k}|<1\right)$ and  the exceptional solution, respectively. So, we have the set
{\small
\begin{equation}
\label{9}
 I=\left\{u_1=\frac{1}{4}(2\lambda_1-\lambda_3),u_2=\frac{1}{4}(2\lambda_2-\lambda_3),u_3=0,u_4=\frac{\lambda_4}{4}\right\},
\end{equation}}
\hspace{-2.622mm}
which contains all the possible valid solutions.
Among the solutions, the smallest value  corresponds to the global minimum of $J_4(\boldsymbol{k})$ (See Appendix~\ref{sec:secA1} for a demonstration).
Let us now consider the different possibilities.
\begin{enumerate}
\item $u_1<u_2,\; u_3,\; u_4$: i.e. the global minimum occurs at $u_1$. In order to have a stable potential, in the strong sense, we impose the condition
\begin{equation}
\label{38aa}
 f(u_1)>0 \quad \Rightarrow\quad \lambda_1>0.
\end{equation}
\item $u_2<u_1,\; u_3,\; u_4$: in this case the strong stability leads to
\begin{equation}
 f(u_2)>0 \quad \Rightarrow\quad \lambda_2>0.
\end{equation}
\item $u_3<u_1,\; u_2,\; u_4$ (remember $u_3=0$): a valid solution requires a positive value for the function (\ref{5}). Let us verify it:
\begin{equation}
\label{40aa}
 f^\prime(0)=\frac{16u_1u_2}{(\lambda_1+\lambda_2-\lambda_3)^2}=\frac{4u_1u_2}{(u_1+u_2)^2}>0.
\end{equation}
 Imposing the strong stability condition
\begin{equation}
\begin{split}
\label{13}
 f(0)&=\frac{\lambda_3^2-4\lambda_1\lambda_2}{4(\lambda_3-\lambda_2-\lambda_1)},\\
&=\frac{4\lambda_1\lambda_2-\lambda_3^2}{8(u_1+u_2)}=\frac{4\lambda_1\lambda_2-\lambda_3^2}{8(u_1+u_2)}>0,
\end{split}
\end{equation}
where $\lambda_3-\lambda_2-\lambda_1=2(u_1+u_2)>0$
 we get
\begin{equation}
 4\lambda_1\lambda_2-\lambda_3^2>0 \quad \rm{or} \quad 4\lambda_1\lambda_2>\lambda_3^2.
\end{equation}
\item $u_4<u_3,\; u_1,\; u_2$ (again $u_3=0$): once more $f^\prime(u_4)$ must be positive. Since each one of the factors $(u_1-u_4),\; (u_2-u_4),\; (u_1+u_2-2u_4)$ are positive, we have
\begin{equation}
 f^\prime(u_4)=\frac{4(u_1-u_4)(u_2-u_4)}{(u_1+u_2-2u_4)^2}>0.
\end{equation}
The strong stability condition produces
\begin{equation}
 f(u_4)=\frac{4\lambda_1\lambda_2-(\lambda_3+\lambda_4)^2}{8(u_1+u_2-2u_4)}>0,
\end{equation}
which means
\begin{equation}
\label{45b}
 4\lambda_1\lambda_2>(\lambda_3+\lambda_4)^2.
\end{equation}
\end{enumerate}
Summarizing, the following are sufficient conditions (but not necessary) to guarantee strong stability of the potential, for all the possible values of the parameters, including the special case $\lambda_1=\lambda_2$:
\begin{subequations}
\label{46b}
\begin{align}
\label{77a}
 \lambda_1&>0,\\
\label{78a}
\lambda_2&>0,\\
\label{20}
4\lambda_1\lambda_2&>\lambda_3^2,\\
\label{21}
4\lambda_1\lambda_2&>(\lambda_3+\lambda_4)^2,
\end{align}
\end{subequations}
where the first two inequalities are also necessary conditions.

%%%%%%%%%%%%%%%%%%%%%%%%%%%%%%%%%%%%%%%%%%%%%%%%%%%%%%%%%%%%%%%%%%%%%%%%%%%%%%%
\subsection{\label{sec:sec322}Global minimum}
According to the general notation introduced in~(\ref{tildepar}), for the economical 3-3-1 model we have
{\small
\begin{equation}
\label{22}
\begin{split}
 \tilde{\boldsymbol{\xi}}&=\begin{pmatrix}
                       \frac{1}{2}(\mu_1^2+\mu_2^2)\\
0\\
0\\
\frac{1}{2}(\mu_1^2-\mu_2^2)
                      \end{pmatrix},\\
&\\[-3mm]
\tilde E&=
\begin{pmatrix}
 \frac{1}{4}(\lambda_1+\lambda_2+\lambda_3)&0&0&\frac{1}{4}(\lambda_1-\lambda_2)\\
0&\frac{\lambda_4}{4}&0&0\\
0&0&\frac{\lambda_4}{4}&0\\
\frac{1}{4}(\lambda_1-\lambda_2)&0&0&  \frac{1}{4}(\lambda_1+\lambda_2-\lambda_3)
\end{pmatrix}.
\end{split}
\end{equation}}
The condition~(\ref{eq-cond}), $\xi_0<|\boldsymbol\xi|$ thus implies that $\mu_1^2+\mu_2^2<|\mu_1^2-\mu_2^2|$. This inequality is fulfilled if
\begin{equation}
\label{2a}
 \mu_1^2, \mu_2^2<0,
\end{equation}
or when at least one of them is negative.

In order to determine the stationary points of the potential $V(\phi_1,\phi_2)$ in Eq.~(\ref{2aaa}) we must solve Eq.~(\ref{eq-stap}):
\begin{equation}
\label{3a}
\begin{split}
&(\tilde E-w\tilde g)\tilde{\boldsymbol{K}}=-\frac{1}{2}\tilde{\boldsymbol\xi} \quad \textrm{with}\quad \tilde{\boldsymbol K}^T\tilde g\tilde {\boldsymbol K}=0 \\
&\left(\textrm{or}\quad  \tilde{\boldsymbol K}^T\tilde g\tilde {\boldsymbol K}> 0\:\: \textrm{when}\:\:  w=0\right) \quad \textrm{and}\quad  K_0>0,
\end{split}
\end{equation}
where $w$ is the Lagrange multiplier. As stated above, for regular values of $w$ with $\det(\tilde E-w\tilde g)\neq 0$ we find solutions to the equation
\[
 \tilde{\boldsymbol{\xi}}^T(\tilde E-w\tilde g)^{-1}\tilde g(\tilde E-w\tilde g)^{-1}\tilde{\boldsymbol\xi}=0,
\]
which gives the following Lagrange multipliers
\begin{equation}
\label{4a}
w_1=\frac{1}{4}\left( \lambda_3-\frac{2\lambda_1\mu_2^2}{\mu_1^2}\right),\quad w_2=\frac{1}{4}\left( \lambda_3-\frac{2\lambda_2\mu_1^2}{\mu_2^2}\right),
\end{equation}
where we have assumed
\begin{equation}
 \mu_1^2\neq 0\quad\rm{and}\quad \mu_2^2\neq0,
\end{equation}
($\mu_i=0$ for $i=1$ or 2 is not relevant as we will show at the end of this section).

The exceptional solutions are obtained from the equation $\det(\tilde E-w\tilde g)=0$, which produces
\begin{equation}
\label{5a}
w_3=-\frac{\lambda_4}{4},\: w_4=\frac{\lambda_3-2\sqrt{\lambda_1\lambda_2}}{4},\: w_5=\frac{\lambda_3+2\sqrt{\lambda_1\lambda_2}}{4}.
\end{equation}
Finally, for the case $\tilde{\boldsymbol K}^T\tilde g\tilde {\boldsymbol K}>0$ we must add the possible solution
\begin{equation}
\label{6a}
 w_6=0.
\end{equation}
Not all $w$  obtained are solutions of Eq.~(\ref{3a}). Let us denote by $\tilde I$ the set of valid solutions which are related to the stationary points of the potential
\begin{equation}
\begin{split}
 \tilde I&=\{\; w \; \textrm{values}\:\:\textrm{in expressions (\ref{4a}), (\ref{5a}) and (\ref{6a})}\\
 & \quad\quad \:\textrm{that are solutions of Eq.~(\ref{3a})}\}.
\end{split}
\end{equation}
The largest $w$  in $\tilde I$ corresponds to the global minimum of the Higgs potential.

%%%%%%%%%%%%%%%%%%%%%%%%%%%%%%%%%%%%%%%%%%%%%%%%%%%%%%%%%%%%%%%%%%%%%%%%%%%%%%%%%%%%%%%%%

\subsubsection{\label{sec:sec323}\hspace{-1mm}Not allowed solutions.}
The global minimum will be among the stationary points in $\tilde I$. By using the Schwarz inequality we can see that the regular and the exceptional solutions, corresponding to the possibility $K_0=|\boldsymbol K|$, implies that the two  scalar triplet vectors at VEV  are linearly dependent, something which does not have any sense (the quantum numbers of the two triplets are different), situation which may be avoided in some cases if only one of the two triplets develops nonzero VEV along its neutral directions. Since at the same time, the global minimum must produce an adequate symmetry breaking pattern (see Appendix~\ref{apendice3}) this kind of solutions are not allowed.
\begin{theorem}
A global minimum with the correct EWSB pattern $SU(3)_L\otimes U(1)_X\rightarrow SU(2)_L\otimes U(1)_Y\rightarrow U(1)_{em}$, where the condition $\xi_0<|\vec{\xi}|$ is required, is given and guaranteed by the stationary points of the classes (Ia) or (IIa) of theorem~\ref{t2} with $K_0>|\vec{K}|$.
\end{theorem}
Let us see this in more detail.

%%%%%%%%%%%%%%%%%%%%%%%%%%%%%%%%%%%%%%%%%%%%%%%%%%%%%%%%%%%%%%%%%%%%%%%%%%%%%%%%%%%%%%%%%%%%

\paragraph{\label{sec:sec331}Regular solutions on the forward light cone.}
\label{sc1}
We start by considering the Lagrange multipliers $w_1$ and $w_2$ in Eq.~(\ref{4a}). Let us define $\textrm{max}\{\tilde I\}$ as the maximum value of the solutions in $\tilde I$. There are two possibilities:
\begin{enumerate}
\item $w_1\in\tilde I$ and $w_1=\textrm{max}\{\tilde I\}$. That is, the point where the global minimum occurs is associated to $w_1$. After solving (\ref{3a}), the global minimum is found at
\begin{equation}
\label{8a}
 \tilde{\boldsymbol{K}}=-\frac{1}{2}(\tilde E-w_1\tilde g)^{-1}\tilde{\boldsymbol\xi}= \begin{pmatrix}
    -\frac{\mu_1^2}{2\lambda_1},&
0,&
0,&
-\frac{\mu_1^2}{2\lambda_1}
   \end{pmatrix}^T,
\end{equation}
under the condition $K_0=-\mu_1^2 / 2\lambda_1>0$. Then we have the equivalence
\begin{equation}
\label{90aa}
 w_1\in\tilde I\quad\Longleftrightarrow\quad \mu_1^2<0.
\end{equation}
Substituting (\ref{8a}) into (\ref{6ab}) we get
\begin{equation}
\label{10a}
\frac{1}{2}(K_01\hspace{-1.5mm}1+K_3\sigma^3)=
\begin{pmatrix}
 -\frac{\mu_1^2}{2\lambda_1} &0\\
0&0
\end{pmatrix}.
\end{equation}
Comparing (\ref{6aa}) and (\ref{10a}) we arrive at the conclusion that no VEV are found in the scalar elements of $\phi_2$, i.e. for this global minimum we have $\langle\phi_2\rangle=0$,  something that should not be accepted, as mentioned above.
\item $w_2\in\tilde I$ and $w_2=\textrm{max}\{\tilde I\}$; in this case the global minimum is associated to $w_2$, and it is found at
\begin{equation}
 \tilde{\boldsymbol{K}}= \begin{pmatrix}
    -\frac{\mu_2^2}{2\lambda_2},&
0,&
0,&
\frac{\mu_2^2}{2\lambda_2}
   \end{pmatrix}^T,
\textrm{ with}\quad K_0=-\frac{\mu_2^2}{2\lambda_2}>0;
\end{equation}
then, for this case we have
\begin{equation}
\label{93aa}
 w_2\in\tilde I \quad \Longleftrightarrow \quad \mu_2^2<0,
\end{equation}
and
\begin{equation}
\label{59b}
\frac{1}{2}(K_01\hspace{-1.5mm}1+K_3\sigma^3)=\begin{pmatrix}
                                                            0&0\\
0&-\frac{\mu_2^2}{2\lambda_2}
                                                           \end{pmatrix},
\end{equation}
implying $\langle\phi_1\rangle=0$ which should not be accepted either.
\end{enumerate}
The two possibilities analyzed above must be discarded because they are unable to implement an adequate symmetry breaking pattern. This conclusion can be expressed in the following way:
\begin{equation}
 \textrm{If}\:\: w_1\:(w_2)\in\tilde I  \Rightarrow  w_1\:(w_2)<\textrm{max}\{\tilde I\}.
\end{equation}

%%%%%%%%%%%%%%%%%%%%%%%%%%%%%%%%%%%%%%%%%%%%%%%%%%%%%%%%%%%%%%%%%%%%%%%%%%%%%%%%%%%%%

\paragraph{\label{sec:sec332}Exceptional solutions on the forward light cone}
The stability condition in~(\ref{20}) implies $w_4<0$; so, according to the discussion following Eq.~(\ref{55a}), $w_4$ cannot give a global minimum either.

In Appendix~(\ref{sec:secA2}) we study and show in detail that the Lagrange multipliers ($w_3$ and $w_5$) do not satisfy the conditions to be global minima either.

We may conclude therefore that
\begin{equation}
\textrm{If}\quad w_3\:(w_4,w_5)\in\tilde I \quad \Rightarrow \quad w_3\:(w_4,w_5)<\textrm{max}\{\tilde I\}.
\end{equation}
%
%%%%%%%%%%%%%%%%%%%%%%%%%%%%%%%%%%%%%%%%%%%%%%%%%%%%%%%%%%%%%%%%%%%%%%%%%%%%%%%%%%%%%%%%%%%%%%%%

\subsubsection{\label{sec:sec34}Allowed solution.}
The only allowed solution to the global minimum lies inside the forward light cone and is associated to the value $w_6=0$, that is
\begin{equation}
\label{96a}
 \textrm{max}\{\tilde I\}=w_6=0.
\end{equation}
From Eq.~(\ref{5a}), the value for $w_3$ allows us to say:
\begin{equation}
\label{68e}
 \textrm{If} \quad w_3\in\tilde I \quad \Rightarrow \quad w_3<0, \quad\textrm{that is}\quad \lambda_4>0.
\end{equation}
Also, from (\ref{20}) and the value for $w_5$ in Eq.~(\ref{5a}) we have that $w_5>0$, implying $w_5>w_6$, which means
\begin{equation}
\label{94}
 w_5\notin \tilde I.
\end{equation}
This result, together with Eqs.~(\ref{138a}) and (\ref{144}) in Appendix~(\ref{sec:secA2}), implies that
\begin{equation}\label{69c}
\sqrt{\lambda_1}\mu_2^2+\sqrt{\lambda_2}\mu_1^2\neq 0.
\end{equation}
The conditions to have the global minimum at $w_6$ require that the solution must satisfy  $-\frac{1}{4}\tilde{\boldsymbol\xi}^T\tilde E^{-1}\tilde g\tilde E^{-1}\tilde{\boldsymbol \xi}<0$, which implies that
\begin{equation}
\label{101a}
 -\frac{64(w_1\mu_1^2)(w_2\mu_2^2)}{(4\lambda_1\lambda_2-\lambda_3^2)^2}<0,
\end{equation}
reproducing the following stationary point:
\begin{equation}
\label{102aa}
 \tilde{\boldsymbol K}=\begin{pmatrix}
                  \frac{4\mu_1^2w_1+4\mu_2^2w_2}{4\lambda_1\lambda_2-\lambda_3^2}\\
0\\
0\\
\frac{4\mu_2^2w_2-4\mu_1^2w_1}{4\lambda_1\lambda_2-\lambda_3^2}
                 \end{pmatrix},
\end{equation}
which is the global minimum as far as
\begin{equation}
\label{102a}
 K_0>0 \quad \Rightarrow \quad 4\mu_1^2w_1+4\mu_2^2w_2>0,
\end{equation}
where the relation~(\ref{20}) has been used.

Using equations (\ref{2a}), (\ref{90aa}), (\ref{93aa}) and~(\ref{96a}), the inequalities in (\ref{101a}) and (\ref{102a}) are fulfilled in the following three different cases (this is going to be seen from another point of view in the following subsection):
\begin{align}
\label{100}
\textrm{Case 1:}\quad w_1,\mu_1^2&<0\quad\textrm{and}\quad w_2,\mu_2^2>0,\\
\textrm{Case 2:}\quad w_1,\mu_1^2&>0\quad\textrm{and}\quad w_2,\mu_2^2<0,\\
\label{16a}
\textrm{Case 3:}\quad w_1,\mu_1^2&<0\quad\textrm{and}\quad w_2,\mu_2^2<0.
\end{align}
A detailed analysis of the three cases shows that only the third one is realistic, and it is the only one consistent with a right implementation of the spontaneous symmetry breaking

%%%%%%%%%%%%%%%%%%%%%%%%%%%%%%%%%%%%%%%%%%%%%%%%

\paragraph{Analysis of case 3.}\label{sd1}
Let us consider the aforementioned Case 3 for which the condition~(\ref{69c}) is immediately satisfied.  The inequalities in Eq.~(\ref{16a}) imply that $\lambda_4>0$ as we are going to see soon:

To prove it, let us assume that $\lambda_4<0,\;\textrm{that is}\; w_3>0$. Since $w_1$ and $w_2$ are negative we have $w_1-w_3<0$ and $w_2-w_3<0$. Then, Eq.~(\ref{51}) in Appendix B is satisfied, but
Eq.~(\ref{55})
becomes $(K_0^2-K_3^2)=\mu_1^2\mu_2^2(w_1-w_3)(w_2-w_3)>0$, which allows for nonzero values in the directions $K_1$ and $K_2$, which in turn implies $w_3(>0)\in\tilde I$, contrary to the conditions expressed in~(\ref{96a}) and~(\ref{68e}). In this development we have used the relation~(\ref{21}) which in turn was used in Eq.~(\ref{b4}). Then, we can claim that
\begin{equation}
\label{108}
 \lambda_4>0.
\end{equation}
This result ($\lambda_4>0$) makes redundant the inequality~(\ref{21}), which may be replaced by the inequality~(\ref{108}).

Now, from (\ref{4a}) and (\ref{16a}) we have that
\begin{equation}
\label{77b}
 \lambda_3<\frac{2\lambda_1\mu_2^2}{\mu_1^2},\quad\textrm{and}\quad
\lambda_3<\frac{2\lambda_2\mu_1^2}{\mu_2^2},
\end{equation}
which does not rule out the possibility of a negative $\lambda_3$ value.

Using the fact that the global minimum occurs at the point given by Eq.~(\ref{102aa}), then from Eqs.~(\ref{6aa}) and (\ref{6ab}), we may claim that
\begin{equation}
\label{79f}
\langle \underline K\rangle=\begin{pmatrix}
\frac{4\mu_2^2w_2}{4\lambda_1\lambda_2-\lambda_3^2}&0\\
0&\frac{4\mu_1^2w_1}{4\lambda_1\lambda_2-\lambda_3^2}
\end{pmatrix},
\end{equation}
where the nonzero VEV must be in both scalar fields, $\phi_1$ and $\phi_2$. Note also in~(\ref{79f}) that the two off-diagonal entries are zero, which implies two things: first the orthogonality condition $\langle\phi_1\rangle^{\textrm{T}}\cdot\langle\phi_2\rangle=0$, and second the electric charge conservation in the model. So, the VEV of the scalars can be written in the following form:
\begin{equation}
\label{111}
 \langle\phi_1\rangle=\frac{1}{\sqrt{2}}\begin{pmatrix}
                       0\\
v_1\\
V_1
                      \end{pmatrix},\quad
\langle\phi_2\rangle=\frac{1}{\sqrt{2}}\begin{pmatrix}
                       v_2\\
0\\
0
                      \end{pmatrix},
\end{equation}
where the inclusion of complex phases does not affect the analysis of the global minimum, as can be seen from the structure of matrix (\ref{79f}).
Note that $\phi_1$ can get VEV at its two neutral directions due to the fact that the minimum state is best achieved in this way, as will be shown at the end of this section; but at this point, the possibility $v_1=0$ or $V_1=0$ is  excluded by this analysis.

Now, using (\ref{6aa}) we have
\begin{align}
\label{22a}
\frac{v_1^2+V_1^2}{2}&=\frac{4\mu_2^2w_2}{4\lambda_1\lambda_2-\lambda_3^2}=\frac{\lambda_3\mu_2^2-2\lambda_2\mu_1^2} {4\lambda_1\lambda_2-\lambda_3^2},\\
\label{23a}
\frac{v_2^2}{2}&=\frac{4\mu_1^2w_1}{4\lambda_1\lambda_2-\lambda_3^2}=\frac{\lambda_3\mu_1^2-2\lambda_1\mu_2^2} {4\lambda_1\lambda_2-\lambda_3^2}.
\end{align}
These equations are equivalent to the tree level constraint equations
\begin{align}
 \mu_1^2+\lambda_1(v_1^2+V_1^2)+\lambda_3\frac{v_2^2}{2}&=0,\\
\mu_2^2+\lambda_3\frac{(v_1^2+V_1^2)}{2}+\lambda_2v_2^2&=0,
\end{align}
the same equations obtained in Refs.~\cite{b3,b4} using a different approach.

At the global minimum the Higgs potential becomes
{\small
\begin{equation}
\label{114}
 V|_{\textrm{\rm{min}.}}=\frac{1}{2}\tilde{\boldsymbol K}^T\tilde{\boldsymbol\xi}=\frac{2\mu_1^2\mu_2^2(w_1+w_2)}{4\lambda_1\lambda_2-\lambda_3}
=\frac{2\mu_1^2\mu_2^2w_2+2\mu_1^2\mu_2^2w_1}{4\lambda_1\lambda_2-\lambda_3^2};
\end{equation}}
using (\ref{29b}), (\ref{22a}) and (\ref{23a}) we get
\begin{equation}
\label{114-1}
 V|_{\textrm{\rm{min}.}}=\frac{\mu_1^2(v_1^2+V_1^2)}{4}+\frac{\mu_2^2v_2^2}{4}<0.
\end{equation}
Therefore, in order to have the deepest minimum value for the potential as stated by Nature, the following conditions are highly suggested:
\begin{align}
\label{86b}
 \mu_1^2<0\quad&\textrm{and}\quad \mu_2^2<0,\\
\label{87b}
v_1,V_1,&v_2\ne0.
\end{align}
These last two expressions explain why Case 3 in~(\ref{16a}) was chosen as the most viable solution. The expression~(\ref{87b}) reveals, for the first time, that the elements of the Higgs scalar triplets develop VEV in all their neutral directions, although a hierarchy among the VEV cannot be concluded from the mathematical point of view.

Finally we must verify the remnant symmetry $U(1)_{\rm{em}}$ left in the scalar potential after the spontaneous symmetry breakdown. For this purpose, we arrange the triplets in Eq.~(\ref{60a}) using the following 2$\times$3 matrix:
\begin{equation}
 \Phi(x)=\begin{pmatrix}
          \phi_1^-&\phi_1^{\prime 0}&\phi_1^0\\
\phi_2^0&\phi_2^{+}&\phi_2^{\prime +}
         \end{pmatrix}.
\end{equation}
An $SU(3)_L\otimes U(1)_X$ gauge transformation $U_G(x)$ maps the scalar triplets as
\begin{equation}
 \phi_i^\alpha\rightarrow \phi_i^{\prime\alpha}=[U_G(x)]^\alpha_\beta\phi_i^\beta,\;\; i=1,2
\end{equation}
Then, the matrix $\Phi(x)$ transforms as
\begin{equation}
\label{135a}
\Phi(x)\rightarrow\Phi^\prime(x)=\Phi(x)U_G^T(x).
\end{equation}
The scalar matrix $\Phi(x)$, in terms of the VEV of the scalar fields, acquires the form
\begin{equation}
 \Phi_{vac}=\begin{pmatrix}
             0&v_1&V_1\\
v_2&0&0
            \end{pmatrix}.
\end{equation}
So, under the transformation (\ref{135a}) we have
\begin{equation}
\label{137a}
 \Phi^{\prime}_{vac}=\Phi_{vac}U_{G}^T.
\end{equation}
Note that the invariance of $\Phi_{vac}$ is always possible for $U_G\ne \mathbbm{1} $, because in (\ref{137a}) we would have more variables than equations.

%%%%%%%%%%%%%%%%%%%%%%%%%%%%%%%%%%%%%%%%%%%%%%%%%%%%%%%%%%%%%%%%%%%%%%%%%%%%%%%%%%%%
%%%%%%%%%%%%%%%%%%%%%%%%%%%%%%%%%%%%%%%%%%%%%%%%%%%%%%%%%%%%%%%%%%%%%%%%%%%%%%%%%%%%

\subsection{\label{sec:sec4}The scalar potential with explicit VEV content.}
An alternative way of writing the scalar potential (\ref{2aaa}), showing explicitly its global minimum is
\begin{equation}
\label{192}
\begin{split}
 V(\phi_1,\phi_2)&=a\left[\phi_1^2+\phi_2^2-\frac{(v_2^2+z^2)}{2}\right]^2+b_1\left(\phi_1^2-\frac{z^2}{2}\right)^2\\
&+ b_2\left(\phi_2^2-\frac{v_2^2}{2}\right)^2+\lambda(\phi_1^\dag\phi_2)
(\phi_2^\dag\phi_1),
\end{split}
\end{equation}
where $z^2=v_1^2+V_1^2$ and $\phi_i^2=\phi_i^\dag\phi_i$,\; $i=1,2$.

This way of writing the scalar potential and the analysis which follows parallels the study used in the first paper of Ref.~\cite{b5} for the THDM; for this reason we may call this form of writing the scalar potential as the Gunion parameterization.

Notice first that $V(\phi_1,\phi_2)$ has six free parameters. A glance to Eq.~(\ref{192}) shows that a sufficient (but not necessary) condition to produce a global minimum at  $\langle\phi_1\rangle=(0,v_1/\sqrt 2,V_1/\sqrt 2)$ and $\langle\phi_2\rangle=(v_2/\sqrt 2,0,0)$ is that
\begin{equation}
\label{122aa}
 a,b_1,b_2,\lambda>0
\end{equation}
(which by the way does not discard the possibility of negative values for some of them since the necessary conditions are $a+b_1>0$ and $a+b_2>0$).

At this point, the criteria for a local minimum becomes
\begin{align}
\label{194}
 \frac{\partial^2V}{(\partial\phi_1^2)^2}&>0 \quad \Rightarrow \quad a+b_1>0,\\
\label{195}
\frac{\partial^2V}{(\partial\phi_2^2)^2}&>0 \quad \Rightarrow \quad a+b_2>0.
\end{align}
and
\begin{equation}
\label{196}
 \det\begin{pmatrix}
\frac{\partial^2V}{(\partial\phi_1^2)^2}&\frac{\partial^2V}{\partial\phi_1^2\partial\phi_2^2}\\
\frac{\partial^2V}{\partial\phi_1^2\partial\phi_2^2}&\frac{\partial^2V}{(\partial\phi_2^2)^2}
     \end{pmatrix}>0 \: \Rightarrow \:(a+b_1)(a+b_2)>a^2.
\end{equation}
On the other hand, comparing (\ref{192}) with (\ref{2aaa}), we see that the parameters in the two representations are related as follows:
\begin{subequations}
\label{93b}
\begin{align}
\label{197}
 \mu_1^2&=-(a+b_1)z^2-av_2^2,\\
\label{198}
\mu_2^2&=-az^2-(a+b_2)v_2^2,\\
\label{128a}
\lambda_1&=a+b_1,\\
\label{129aa}
\lambda_2&=a+b_2,\\
\label{130aa}
\lambda_3&=2a,\\
\label{131aa}
\lambda_4&=\lambda,
\end{align}
\end{subequations}
such that the relations (\ref{194}), (\ref{195}) and (\ref{196}) correspond to the inequalities (\ref{77a}), (\ref{78a}) and (\ref{20}), respectively.

Examining now Eqs.~(\ref{197}) and (\ref{198}) we have
\begin{align}
 \mu_1^2&=-(a+b_1)z^2-av_2^2=-\lambda_1z^2-\frac{\lambda_3}{2}v_2^2,\\
\mu_2^2&=-az^2-(a+b_2)v_2^2=-\frac{\lambda_3}{2}-\lambda_2v_2^2,
\end{align}
which can be written as
\begin{equation}
 \begin{pmatrix}
  \mu_1^2\\
\mu_2^2
 \end{pmatrix}
=\begin{pmatrix}
  -\lambda_1&-\frac{\lambda_3}2\\
-\frac{\lambda_3}{2}&-\lambda_2
 \end{pmatrix}
\begin{pmatrix}
 z^2\\
v_2^2
\end{pmatrix}.
\end{equation}
Solving, we obtain
\begin{equation}
\label{138aaa}
 \frac{1}{2}\begin{pmatrix}
             z^2\\
v_2^2
            \end{pmatrix}
=
\begin{pmatrix}
 \frac{\lambda_3\mu_2^2-2\lambda_2\mu_1^2}{4\lambda_1\lambda_2-\lambda_3^2}\\
\frac{\lambda_3\mu_1^2-2\lambda_1\mu_2^2}{4\lambda_1\lambda_2-\lambda_3^2}
\end{pmatrix}=
\begin{pmatrix}
 \frac{4\mu_2^2w_2}{4\lambda_1\lambda_2-\lambda_3^2}\\
\frac{4\mu_1^2w_1}{4\lambda_1\lambda_2-\lambda_3^2}
\end{pmatrix}.
\end{equation}
The fact that $z^2>0$ and $v_2^2>0$ implies that the following product must remain always positive:
\begin{equation}
\mu_2^2\: w_2>0\quad{\rm and}\quad \mu_1^2\: w_1>0,
\end{equation}
which shows in a different way the validity of the classification introduced in (\ref{100})-(\ref{16a}) for the required symmetry breaking.

%%%%%%%%%%%%%%%%%%%%%%%%%%%%%%%%%%%%%%%%%%%%%%%%%%%%%%%%%%%%%%%%%%%%%%%%%%%%%%%%%%%%

\section{\label{sec:sec5}New parameterizations}
The search and study of possible new parametrizations give us the possibility of checking some of the previously obtained results.
New parameterizations for the invariant scalar products, different to  the ones given in (\ref{8}), can be constructed. We will partially study two cases and, for each one, we will verify the symmetry breaking $SU(3)_L\otimes U(1)_X\rightarrow U(1)_{\rm{em}}$ following the analysis of Sect.~\ref{sec:sec22}.

A new parameterization for the scalar potential (\ref{2aaa}) is obtained by defining the variables
\begin{equation}
 K_1=\phi_1^\dag\phi_1,\quad K_2=\phi_2^\dag\phi_2,\quad K_3=\phi_1^\dag\phi_2,
\end{equation}
so that the potential is written as
\begin{equation}
\label{120}
\begin{split}
 V&=\mu_1^2K_1+\mu_2^2K_2+\lambda_1K_1^2+\lambda_2K_2^2+\lambda_3K_1K_2+\lambda_4K_3K_3^*,\\
V&=\tilde{\boldsymbol K}\cdot\tilde{\boldsymbol \xi}+\tilde{\boldsymbol K}\cdot\tilde E\cdot\tilde{\boldsymbol K},
\end{split}
\end{equation}
with
{\small
\begin{equation}
 \tilde{\boldsymbol K}=
\begin{pmatrix}
 K_1\\
K_2\\
K_3\\
K_3^*
\end{pmatrix},
\:
\tilde{\boldsymbol \xi}=
\begin{pmatrix}
 \mu_1^2\\
\mu_2^2\\
0\\
0
\end{pmatrix},\:
\tilde E=
\begin{pmatrix}
 \lambda_1&\frac{\lambda_3}{2}&0&0\\
\frac{\lambda_3}{2}&\lambda_2&0&0\\
0&0&0&\frac{\lambda_4}{2}\\
0&0&\frac{\lambda_4}{2}&0
\end{pmatrix}.
\end{equation}}
The new parameters satisfy the constraints
\begin{align}
 K_1&\ge0\\
K_2&\ge0\\
\label{124}
K_1K_2&\ge K_3K_3^*,\;\;{\mbox {\rm or}}\;\; \tilde{\boldsymbol K}\cdot \tilde g\cdot
\tilde{\boldsymbol K}\ge 0,
\end{align}
with
\begin {equation}
 \tilde g=\begin{pmatrix}
    0&1/2&0&0\\
1/2&0&0&0\\
0&0&0&-1/2\\
0&0&-1/2&0
   \end{pmatrix},
\end {equation}
where (\ref{124}) comes from the Schwarz inequality.

Now, for the case $K_1K_2>K_3K_3^*$, we calculate the stationary point of the potential (\ref{120}). To do this  we solve the equation $\tilde E\cdot\tilde{\boldsymbol K}=-\frac{1}{2}\tilde{\boldsymbol \xi}$, and we get
\begin{equation}
 \tilde{\boldsymbol K}=
\begin{pmatrix}
                    \frac{\lambda_3\mu_2^2-2\lambda_2\mu_1^2}{4\lambda_1\lambda_2-\lambda_3^2}\\
\frac{\lambda_3\mu_1^2-2\lambda_1\mu_2^2}{4\lambda_1\lambda_2-\lambda_3^2}\\
0\\
0
                   \end{pmatrix},
\end{equation}
which coincides with the results in (\ref{22a}) and (\ref{23a}).

To obtain another different parameterization, let us construct the following $SU(3)_L\otimes U(1)_X$ gauge invariant array
\begin{equation}
\label{127a}
\underline K=\begin{pmatrix}
              (\phi_1^\dag\phi_1)^2&(\phi_2^\dag\phi_1)(\phi_1^\dag\phi_2)\\
(\phi_2^\dag\phi_1)(\phi_1^\dag\phi_2)&(\phi_2^\dag\phi_2)^2
             \end{pmatrix}.
\end{equation}
This matrix is real, symmetric and positive. We now write this matrix using the basis
\[\begin{pmatrix}
1&0\\
0&1
\end{pmatrix},\begin{pmatrix}
1&0\\
0&-1
\end{pmatrix},\begin{pmatrix}
0&1\\
1&0
\end{pmatrix};\]
that is
\begin{equation}
\underline K=K_1 \begin{pmatrix}
1&0\\
0&1
\end{pmatrix}+K_2\begin{pmatrix}
1&0\\
0&-1
\end{pmatrix}+K_3\begin{pmatrix}
0&1\\
1&0
\end{pmatrix},
\end{equation}
which, compared with (\ref{127a}) gives
\begin{align}
\label{129a}
 (\phi_1^\dag\phi_1)^2&=K_1+K_2,\: \Longrightarrow \: \phi_1^\dag\phi_1=\sqrt{K_1+K_2},\\
\label{130a}
(\phi_2^\dag\phi_2)^2&=K_1-K_2,\: \Longrightarrow \: \phi_2^\dag\phi_2=\sqrt{K_1-K_2},\\
\label{131a}
(\phi_2^\dag\phi_1)&(\phi_1^\dag\phi_2)=K_3.
\end{align}
Due to the positivity of (\ref{127a}) we have
\begin{equation}
 K_1\ge0,\quad K_1^2-K_2^2-K_3^2\ge0.
\end{equation}
Notice however that the scalar potential is not a polynomial function of the parameters $K_1,\; K_2$ and $K_3$ [see the relations~(\ref{129a}) and~(\ref{130a})].

%%%%%%%%%%%%%%%%%%%%%%%%%%%%%%%%%%%%%%%%%%%%%%%%%%%%%%%%%%%%%%%%%%%%%%%%%%%%%%%%%%%%

\section{\label{sec:sec6}The potential after the electroweak symmetry breaking}
To analyze the form of the scalar potential after the electroweak symmetry has been broken, we may throw some insight into the physical problem, as we are now going to see.
We start by assuming a stable potential which leads to the desired symmetry breaking pattern as discussed in the previous sections, and thus we see what the consequences are for the resulting physical fields. For this purpose we work in the unitary gauge and use a basis for the scalar fields such that the VEV in~(\ref{111}) hold. Furthermore, the relation
\begin{align}
 \rm{Im}&\:\phi_1^{\prime 0}=0
\end{align}
immediately produces one Goldstone boson ($Go_1$) which is eaten up by one of the CP-odd gauge bosons.

We use as usual the following shifted Higgs fields in the two triplets
\begin{equation}
 \phi_1=\frac{1}{\sqrt{2}}\begin{pmatrix}
                           \sqrt{2}\phi_1^-\\
v_1+H_1^\prime\\
V_1+H_1+iA_1
                          \end{pmatrix},
\phi_2=\frac{1}{\sqrt{2}}
\begin{pmatrix}
 v_2+H_2+iA_2\\
\sqrt{2}\phi_2^+\\
\sqrt{2}\phi_2^{\prime +}
\end{pmatrix}.
\end{equation}%}
We may now proceed to find the remaining Goldstone bosons and the physical Higgs fields (three CP-even and one CP-odd).

It is convenient to decompose $\tilde{\boldsymbol K}$ according to the power of the physical fields
\begin{equation}
\label{126}
 \tilde{\boldsymbol K}=\tilde{\boldsymbol K}_{\{0\}}+\tilde{\boldsymbol K}_{\{1\}}+\tilde{\boldsymbol K}_{\{2\}},
\end{equation}
with
%
%\begin{widetext}
%\begin{onecolumn}
%{\onecolumn
{\small
\begin{align}
\label{127}
 \tilde{\boldsymbol K}_{\{0\}}&=\begin{pmatrix}
                        \frac{V_1^2}{2}+\frac{v_1^2}{2}+\frac{v_2^2}{2}\\
0\\
0\\
 \frac{V_1^2}{2}+\frac{v_1^2}{2}-\frac{v_2^2}{2}
                       \end{pmatrix},\\%\quad
\tilde{\boldsymbol K}_{\{1\}}&=\begin{pmatrix}
                           V_1H_1+v_2H_2+v_1H_1^\prime\\
\frac{V_1}{\sqrt{2}}\phi_2^{\prime -}+\frac{V_1}{\sqrt{2}}\phi_2^{\prime +}+\frac{v_1}{\sqrt{2}}\phi_2^{-}+\frac{v_1}{\sqrt{2}}\phi_2^{+}
+\frac{v_2}{\sqrt{2}}\phi_1^{ -}+\frac{v_2}{\sqrt{2}}\phi_1^{+}\\
\frac{\rm{i}V_1}{\sqrt{2}}\phi_2^{\prime -}-\frac{\rm{i}V_1}{\sqrt{2}}\phi_2^{\prime +}+\frac{\rm{i}v_1}{\sqrt{2}}\phi_2^{-}-
\frac{\rm{i}v_1}{\sqrt{2}}\phi_2^{+}+\frac{\rm{i}v_2}{\sqrt{2}}\phi_1^{-}-\frac{\rm{i}v_2}{\sqrt{2}}\phi_1^{+}\\
V_1H_1+v_1H_1^\prime-v_2H_2
                          \end{pmatrix},
\end{align}}
{\scriptsize
\begin{equation}
\begin{split}
\label{128}
\tilde{\boldsymbol K}_{\{2\}}=
\frac{1}{2}
\begin{pmatrix}
 H_2^2+H_1^2+A_2^2+A_1^2+H_1^{\prime 2}+2\phi_2^{\prime +}\phi_2^{\prime-}
+2\phi_2^+\phi_2^-+2\phi_1^+\phi_1^-\\
\\[-1.5mm]
\sqrt{2}\phi_2^{\prime -}H_1+\sqrt{2}\phi_2^{\prime +}H_1+\rm i\sqrt{2}\phi_2^{\prime -}A_1-\rm i\sqrt{2}\phi_2^{\prime +}A_1+\sqrt{2}\phi_2^{-}H_1^\prime+\\
\sqrt{2}\phi_2^{+}H_1^\prime+\sqrt{2}\phi_1^{-}H_2+\sqrt{2}\phi_1^{+}H_2
-\rm i\sqrt{2}\phi_1^-A_2+\rm i\sqrt{2}\phi_1^+A_2
\\
\\[-1.5mm]
\rm i\sqrt{2}\phi_2^{\prime -}H_1-\rm i\sqrt{2}\phi_2^{\prime +}H_1-\sqrt{2}\phi_2^{\prime -}A_1-\sqrt{2}\phi_2^{\prime +}A_1+\rm i\sqrt{2}\phi_2^{-}H_1^\prime-\\
\rm i\sqrt{2}\phi_2^{+}H_1^\prime+\rm i\sqrt{2}\phi_1^{-}H_2-\rm i\sqrt{2}\phi_1^{+}H_2+\sqrt{2}\phi_1^-A_2+\sqrt{2}\phi_1^+A_2\\
\\[-1.5mm]
H_1^2+A_1^2+H_1^{\prime 2}-H_2^2-A_2^2-2\phi_2^{\prime +}\phi_2^{\prime -}-2\phi_2^+\phi_2^--2\phi_1^+\phi_1^-
\end{pmatrix}.
\end{split}
\end{equation}}
%\end{onecolumn}
%\twocolumn
%\end{widetext}
%
The global minimum of the potential occurs when $w=0$ in Eq.~(\ref{3a}). This leads to
\begin{equation}
\label{129}
 \tilde E\tilde{\boldsymbol K}_{\{0\}}=-\frac{1}{2}\tilde{\boldsymbol \xi}.
\end{equation}
Using equations
(\ref{126}) to (\ref{129}), we get for the potential in Eq.~(\ref{eq-vtil})
\begin{equation}
 V=V_{\{0\}}+V_{\{2\}}+V_{\{3\}}+V_{\{4\}},
\end{equation}
where $V_{\{k\}}$ are the terms of order $k^{\rm th}$ in the physical fields
\begin{align}
V_{\{0\}}&=\frac{1}{2}\tilde{\boldsymbol K}_{\{0\}}\cdot\tilde{\boldsymbol\xi},\\
\label{134d}
V_{\{2\}}&=\tilde{\boldsymbol K}_{\{1\}}\cdot\tilde E\cdot\tilde{\boldsymbol K}_{\{1\}},\\
V_{\{3\}}&=2\tilde{\boldsymbol K}_{\{1\}}\cdot\tilde E\cdot\tilde{\boldsymbol K}_{\{2\}},\\
V_{\{4\}}&=\tilde{\boldsymbol K}_{\{1\}} \cdot\tilde E\cdot\tilde{\boldsymbol K}_{\{2\}}.
\end{align}
The second order terms~(\ref{134d}) determine the masses of the physical Higgs fields and the remaining Goldstone bosons
\begin{equation}
\label{136b}
\begin{split}
V_{\{2\}}&=\frac{1}{2}
 \begin{pmatrix}
  H_1&H_2&H_1^\prime
 \end{pmatrix}
{\cal M}_{\rm{neutral}}^2
\begin{pmatrix}
 H_1\\
H_2\\
H_1^\prime
\end{pmatrix}\\
&+
\begin{pmatrix}
 \phi_1^+&\phi_2^+&\phi_2^{\prime +}
\end{pmatrix}
{\cal M}_{\rm{charged}}^2
\begin{pmatrix}
\phi_1^-\\ \phi_2^-\\ \phi_2^{\prime -}
\end{pmatrix},
\end{split}
\end{equation}
with
\begin{align}
\label{138d}
{\cal M}_{\rm{neutral}}^2&=
\begin{pmatrix}
 2\lambda_1V_1^2&\lambda_3v_2V_1&2\lambda_1v_1V_1\\
\lambda_3v_2V_1&2\lambda_2v_2^2&\lambda_3v_1v_2\\
2\lambda_1v_1V_1&\lambda_3v_1v_2&2\lambda_1v_1^2
\end{pmatrix},\\
\label{139d}
{\cal M}_{\rm{charged}}^2&=\frac{\lambda_4}{2}
\begin{pmatrix}
 v_2^2&v_1v_2&v_2V_1\\
v_1v_2&v_1^2&v_1V_1\\
v_2V_1&v_1V_1&V_1^2
\end{pmatrix}.
\end{align}
Clearly, the fields $A_1$ and $A_2$ are massless, providing two other CP-odd Goldstone bosons $Go_2$ and $Go_3$.
The neutral sector~(\ref{138d}) provides a CP-even Goldstone boson $Ge_4$ and two CP-even massive scalars $Hgg_1$ and $Hgg_2$ with masses
{\small
\begin{equation}
\begin{split}
&M^2_{Hgg_1,Hgg_2}=(v_1^2+V_1^2)\lambda_1+v_2^2\lambda_2\\
&\pm\sqrt{[(v_1^2+V_1^2)\lambda_1+v_2^2\lambda_2]^2+v_2^2(v_1^2+V_1^2)
(\lambda_3^2-4\lambda_1\lambda_2)}.
\end{split}
\end{equation}}
Now, the stability of the potential requires that $\lambda_1>0$, $\lambda_2>0$ and $4\lambda_1\lambda_2>\lambda_3^2$ (see Eqs.~(\ref{77a}), (\ref{78a}) and (\ref{20})), which in turn implies a positive value for the former masses of the scalar fields predicted by the model.

For the charged sector~(\ref{139d}) we get two zero eigenvalues corresponding to four Goldstone bosons $G_5^\pm,G_6^\pm$, two CP-even and two CP-odd, and two charged scalars, one CP-even and one CP-odd, with a degenerate mass $\frac{\lambda_4}{2}(v_1^2+v_2^2+V_1^2)$, which, according with Eq.~(\ref{108}), is positive.

The former analysis is in agreement with the results obtained in Refs.~\cite{b3} and \cite{b4}.

%%%%%%%%%%%%%%%%%%%%%%%%%%%%%%%%%%%%%%%%%%%%%%%%%%%%%%%%%%%%%%%%%%%%%%%%%%%%%%%

\section{\label{sec:sec7}Conclusions}
A detailed study of the scalar potential for the economical 3-3-1 model has been carried through. In order to have an acceptable theory, this potential should be stable; that is, it should be bounded from below and lead to the correct EWSB pattern observed in Nature.

For the scalar potential as presented in Eq.~(\ref{2aaa}), the following are the conditions which guarantee strong stability:
\begin{enumerate}
\item Necessary and sufficient conditions:
\begin{equation*}
\lambda_1>0\;\;\;{\mbox {\rm and}}\;\;\;\lambda_2>0.
\end{equation*}
\item Sufficient (but not necessary) conditions
\begin{equation*}
4\lambda_1\lambda_2>\lambda_3^2 \;\;{\mbox {\rm and}}\;\;\; 4\lambda_1\lambda_2>(\lambda_3+\lambda_4)^2.
\end{equation*}
\end{enumerate}
Now, at the global minimum of the potential, $\lambda_4>0$ is required; a condition which makes redundant the last inequality. And the inequality $4\lambda_1\lambda_2>\lambda_3^2$ is a necessary condition in order that the square mass for the physical Higgs be positive.

Additional constraints coming from our analysis are:
\begin{itemize}
\item The criteria used to find the minimum state leads us to assure that the second order coefficients  $\mu_1^2$ and $\mu_2^2$ in the scalar potential must be negative.

\item The required EWSB allows us to conclude that both scalar triplets must develop nonzero VEV. Additionally, the VEV  are found to be necessary along the three electrically neutral directions of the scalar fields.

\item In the main text, specific new relations among several parameters of the scalar potential were derived, as for example that $\sqrt{\lambda_1}\mu_2^2+\sqrt{\lambda_2}\mu_1^2\neq 0$. This condition is related to the existence of a \textit{critical point} on the scalar potential.

\item  In Refs. \cite{b3,b4} $\lambda_3$ was declared as a negative value parameter. Here we have shown that under special circumstances it can take positive values, constrained by
\[\lambda_3<min\{2\lambda_1|\mu_2^2/\mu_1^2|,\;\; 2\lambda_2|\mu_1^2/\mu_2^2|\}\]
\item Unfortunately, from the mathematical point of view we could not establish a hierarchy among $V_1,\; v_1$ and $v_2$, unless a fine tuning is introduced (from the physical point of view we know that $V_1>>v_2>>v_1$\cite{b4}).
\end{itemize}
But the most important conclusion of our study is that the conditions for strong stability of the scalar potential, guarantee positive masses for the scalar fields predicted by the model. This outstanding result shows the consistency of the economical 3-3-1 model, something that should not be taken for granted due to the scarce number of parameters to deal with.

Notice that the inclusion of imaginary VEV do not alter the minimum of the scalar potential, due to the fact that $\langle\phi_1\rangle^T\cdot\langle\phi_2\rangle=0$ in Eq.~(\ref{79f}).

Notice also that in order to implement the mathematical method in this particular model, the criteria for stability were straightened, with a new theorem  proved in Appendix \ref{sec:secA1}.

The mathematical analysis presented here may be extended to other 3-3-1 models with three or more Higgs scalar triplets (work in progress). For these other models the Gunion parameterization may not be implemented easily.

The parameterization given in Sect.~\ref{sec:sec2} for the scalars, using orbital variables, is not unique. Other acceptable parameterizations can be found in Sect.~\ref{sec:sec5}. These new schemes seem to work well and  deserve more attention, in particular the last parameterization used has the additional property that the scalar product terms are $SU(3)_L\otimes U(1)_X$ gauge invariant.

Finally we want to mention that some results presented here, either coincide or are compatible with partial results already published in Refs.~\cite{b3,b4}.

%%%%%%%%%%%%%%%%%%%%%%%%%%%%%%%%%%%%%%%%%%%%%%%%%%%%%%%%%%%%%%%%%%%%%%%%%%%%%%%%

\section*{ACKNOWLEDGEMENTS}
%
%\begin{acknowledgement}
We thank C.Garcia-Canal for calling our attention to Refs.~\cite{b1} and \cite{b2}, and to D. Gomez-Dumm for explaining us the relevance of the scalar potential introduced in~(\ref{192}). L.A.S. acknowledges partial financial support from DIME at Universidad Nacional de Colombia, Sede Medell\'\i n.
%\end{acknowledgement}

\appendix

\section{\label{sec:secA1}The smallest Lagrange multiplier as the global minimum of the function $J_4(\boldsymbol{k})$}
Let $\boldsymbol{p}$ and $\boldsymbol{q}$ be two stationary points with Lagrange multipliers $u_p$ and $u_q$ respectively, with $|\boldsymbol{p}|=|\boldsymbol{q}|=1$ (we will consider later the case $u_p=0$, $|\boldsymbol{p}|< 1$). Both
$\boldsymbol{p}$ and $\boldsymbol{q}$ must satisfy
\begin{equation}
\label{48}
 (E-u_p)\boldsymbol{p}=-\boldsymbol\eta\quad\mathrm {and}\quad (E-u_q)\boldsymbol{q}=-\boldsymbol\eta.
\end{equation}
At these two stationary points, $J_4(\boldsymbol{k})$ takes the values
\begin{align}
J_4(\boldsymbol{p})&=\eta_{00}+u_p+\boldsymbol \eta^T\cdot \boldsymbol{p},\\
J_4(\boldsymbol{q})&=\eta_{00}+u_q+\boldsymbol \eta^T\cdot \boldsymbol{q},
\end{align}
where we have used Eqs.~(\ref{eq-flam}),~(\ref{11d}) and~(\ref{48}).
Subtracting we obtain
\begin{equation}
\label{51aa}
 J_4(\boldsymbol{p})-J_4(\boldsymbol{q})=u_p-u_q+\boldsymbol\eta^T\cdot(\boldsymbol{p}-\boldsymbol{q}).
\end{equation}
Now, recalling that $(E-u_p)^T=E-u_p$, we transpose Eqs.~(\ref{48})
\begin{equation}
  \boldsymbol{p}^T(E-u_p)=-\boldsymbol\eta^T,\qquad
 \boldsymbol{q}^T(E-u_q)=-\boldsymbol\eta^T.
\end{equation}
Multiplying by $\boldsymbol{q}$ and $\boldsymbol{p}$, we have
\begin{align}
\label{53}
 \boldsymbol{p}^T\cdot (E-u_p)\boldsymbol{q}=-\boldsymbol\eta^T\cdot\boldsymbol{q},\\
\label{54}
\boldsymbol{q}^T\cdot (E-u_q)\boldsymbol{p}=-\boldsymbol\eta^T\cdot\boldsymbol{p}.
\end{align}
Subtracting Eqs.~(\ref{53}) and (\ref{54}) it is obtained that
\begin{equation}
 (u_q-u_p)\boldsymbol{p}^T\cdot\boldsymbol{q}=\boldsymbol{\eta}^{\mathrm{T}}\cdot (\boldsymbol{p}-\boldsymbol{q}),
\end{equation}
which we place into (\ref{51aa}) to finally obtain
\begin{equation}
\label{56a}
\begin{split}
 J_4(\boldsymbol{p})-J_4(\boldsymbol{q})&=u_p-u_q+(u_q-u_p)\boldsymbol{p}^T\cdot\boldsymbol{q},\\
&=(u_p-u_q)(1-\boldsymbol{p}^T\cdot\boldsymbol{q}),
\end{split}
\end{equation}
where $\boldsymbol{p}^T\cdot\boldsymbol{q}=|\boldsymbol{p}||\boldsymbol{q}|$ $\cos\theta =\cos\theta < 1$ \footnote{If $\vert \boldsymbol{p}\vert <1$, then $\boldsymbol{p}^T \boldsymbol{q}=\vert \boldsymbol{p}\vert \vert \boldsymbol{q}\vert$ $\cos\theta < 1$}.
Notice that $\cos\theta$ cannot be equal to $1$, because $\boldsymbol{p}$ and $\boldsymbol{q}$ cannot be parallel: if we assume that they are  parallel to each other, Eq.~(\ref{48}) leads to
\begin{equation}
 (u_p-u_q)\boldsymbol{p}=0, \quad \textrm{and then}\quad u_p=u_q,
\end{equation}
but we have assumed $u_p\neq u_q$. So, in all the cases we would have
\begin{equation}
 (1-\boldsymbol{p}^T\cdot\boldsymbol{q})>0.
\end{equation}
From Eq.~(\ref{56a}), we finally conclude that
\begin{equation}
 \textrm{if}\quad u_p<u_q\Leftrightarrow J_4(\boldsymbol{p})<J_4(\boldsymbol{q}).\quad \quad \qed
\end{equation}
%
%%%%%%%%%%%%%%%%%%%%%%%%%%%%%%%%%%%%%%%%%%%%%%%%%%%%%%%%%%%%%%%%%%%%%%%%%%%%%%%%%%

%%%%%%%%%%%%%%%%%%%%%%%%%%%%%%%%%%%%%%%%%%%%%%%%%%%%%%%%%%%%%%%%%%%%%%%%%%%%%%%%%%

\section{\label{sec:secA2}The exceptional solutions $w_3$ and $w_5$}
In what follows we are going to find the conditions which avoid that the Lagrange multipliers $w_3=-\frac{\lambda_4}{4}$ and $w_5=\frac{\lambda_3+2\sqrt{\lambda_1\lambda_2}}{4}$ be global minima.

%%%%%%%%%%%%%%%%

\subsection{The exceptional solution $w_3$:}

Let us assume that $w_3$ is the largest value among the acceptable solutions in $\tilde I$, that is
$ w_3=\textrm{max}\{\tilde I\}.$
For $w_3$, let us solve the equation $(\tilde E-w_3\tilde g)\tilde{\boldsymbol K}=-\frac{1}{2}\tilde{\boldsymbol\xi}$, where
\begin{equation}
 \tilde E-w_3\tilde g=\begin{pmatrix}
                       \frac{\lambda_1+\lambda_2+\lambda_3+\lambda_4}{4}&0&0&\frac{\lambda_1-\lambda_2}{4}\\
0&0&0&0\\
0&0&0&0\\
\frac{\lambda_1-\lambda_2}{4}&0&0&\frac{\lambda_1+\lambda_2-\lambda_3-\lambda_4}{4}
                      \end{pmatrix}.
\end{equation}
By looking the parameters in Eq.~(\ref{22}), we see that the orbital variables $K_1$ and $K_2$ would be arbitrary. But by the use of Eq.~(\ref{6aa}) the cases $K_1\neq0$ or $K_2\neq0$ imply that $\phi_1^\dag\phi_2 \neq0$, i.e. we would have electric charge breaking.

If $K_1=0$ and $K_2=0$, we focus on the variables $K_0$ and $K_3$:
{\small
\begin{equation}
 \begin{pmatrix}
  \frac{\lambda_1+\lambda_2+\lambda_3+\lambda_4}{4}&\frac{\lambda_1-\lambda_2}{4}\\
\frac{\lambda_1-\lambda_2}{4}&\frac{\lambda_1+\lambda_2-\lambda_3-\lambda_4}{4}
 \end{pmatrix}
\begin{pmatrix}
 K_0\\
K_3
\end{pmatrix}=-\frac{1}{4}\begin{pmatrix}
\mu_1^2+\mu_2^2\\
\mu_1^2-\mu_2^2
\end{pmatrix},
\end{equation}}
then
{\small
\begin{equation}
\label{b4}
\begin{pmatrix}
 K_0\\
K_3
\end{pmatrix}=a
\begin{pmatrix}
 (-2\lambda_1+\lambda_3+\lambda_4)\mu_2^2+(-2\lambda_2+\lambda_3+\lambda_4)\mu_1^2\\
-(2\lambda_2+\lambda_3+\lambda_4)\mu_1^2+(2\lambda_1+\lambda_3+\lambda_4)\mu_2^2
\end{pmatrix},
\end{equation}
}
with $a=1/\left(4\lambda_1\lambda_2-(\lambda_3+\lambda_4)^2\right)$.
The global minimum requires that
\begin{equation}
\label{51}
\begin{split}
 K_0&>0 \: \Rightarrow \: (-2\lambda_1+\lambda_3+\lambda_4)\mu_2^2+(-2\lambda_2+\lambda_3+\lambda_4)\mu_1^2=\\
&(-2\lambda_1\mu_2^2+\lambda_4\mu_1^2
+\lambda_3\mu_1^2)+(-2\lambda_2\mu_1^2+\lambda_4\mu_2^2
+\lambda_3\mu_2^2)=\\
&4\mu_1^2(w_1-w_3)+4\mu_2^2(w_2-w_3)>0,
\end{split}
\end{equation}
and
\begin{equation}
\label{55}
\begin{split}
 &K_0^2-K_3^2=0  \Rightarrow \\
& (2\lambda_1\mu_2^2-\lambda_3\mu_1^2-\lambda_4\mu_1^2)(2\lambda_2\mu_1^2-\lambda_3\mu_2^2-\lambda_4\mu_2^2)=\\
&\mu_1^2\mu_2^2(w_1-w_3)(w_2-w_3)=0,
\end{split}
\end{equation}
which implies either $w_3=w_1$ or $w_3=w_2$. These solutions were already studied  in Sect.~\ref{sec:sec331}.

%%%%%%%%%%%%%%%%%%%%%%%%%%%%%%%%%%%%%%%%%%%%%%%%%%%%%%%%%%%%%%%%%%%%%%%%%%%%%

\subsection{The exceptional solution $w_5$:}

In this case we solve the equation $(\tilde E-w_5\tilde g)\tilde{\boldsymbol K}=-\frac{1}{2}\tilde{\boldsymbol\xi}$, where  the matrix $\tilde E-w_5\tilde g$ is equal to
{\scriptsize
\begin{equation}
\begin{pmatrix}
\frac{\lambda_1+\lambda_2-2\*\sqrt{\lambda_1\*\lambda_2}}{4}&0&0&\frac{\lambda_1-\lambda_2}{4}\\
0&\frac{\lambda_3+\lambda_4+2\sqrt{\lambda_1\lambda_2}}{4}&0&0\\
0&0&\frac{\lambda_3+\lambda_4+2\sqrt{\lambda_1\lambda_2}}{4}&0\\
\frac{\lambda_1-\lambda_2}{4}&0&0&\frac{\lambda_1+\lambda_2+2\*\sqrt{\lambda_1\*\lambda_2}}{4}
\end{pmatrix}.
\end{equation}}
From (\ref{21}) we have $\frac{\lambda_3+\lambda_4+2\sqrt{\lambda_1\lambda_2}}{4}>0$, then $K_1=K_2=0$. The equation relating $K_0$ and $K_3$ is
{\small
\begin{equation}
\label{59}
 \frac{1}{4}\begin{pmatrix}
             \left(\sqrt{\lambda_1}-\sqrt{\lambda_2}\right)^2&\lambda_1-\lambda_2\\
\lambda_1-\lambda_2&\left(\sqrt{\lambda_1}+\sqrt{\lambda_2}\right)^2
            \end{pmatrix}
\begin{pmatrix}
 K_0\\
K_3
\end{pmatrix}=-\frac{1}{4}
\begin{pmatrix}
 \mu_1^2+\mu_2^2\\
\mu_1^2-\mu_2^2
\end{pmatrix}.
\end{equation}}
Notice that the $2\times 2$ matrix in the left hand side of~(\ref{59}) is not invertible. Its entries are therefore linearly dependent
{\small
\begin{align}
\label{135}
 \left(\sqrt{\lambda_1}-\sqrt{\lambda_2}\right)^2K_0+(\lambda_1-\lambda_2)K_3&=-(\mu_1^2+\mu_2^2),\\
\label{136}
(\lambda_1-\lambda_2)K_0+\left(\sqrt{\lambda_1}+\sqrt{\lambda_2}\right)^2K_3&=-(\mu_1^2-\mu_2^2).
\end{align}
}
We will solve these equations in the following two cases:
\begin{description}

 \item[i)] $\lambda_1=\lambda_2$: then, from (\ref{135}), we have
\begin{equation}
\label{138a}
\mu_1^2+\mu_2^2=0,
\end{equation}
which together with (\ref{136}), gives
\begin{equation}
K_3=-\frac{\mu_1^2}{2\lambda_1}.
 \end{equation}
Additionally
\begin{equation}
\label{138}
 K_0^2-K_3^2=0;
\end{equation}
then
\begin{equation}
\label{139}
 K_0=\pm K_3.
\end{equation}
In both  cases
\begin{equation}
\label{140}
 \underline K=\begin{pmatrix}
      0&0\\
0&\frac{K_0+K_3}{2}
     \end{pmatrix}\quad\textrm{or}\quad
 \underline K=\begin{pmatrix}
      \frac{K_0+K_3}{2}&0\\
0&0
     \end{pmatrix}.
\end{equation}
\item[ii)] $\lambda_1\neq\lambda_2$: in this case, taking into account Eqs.~(\ref{135}) and (\ref{136}), the entries in the right hand side of (\ref{59}) must be such that
\begin{equation}\label{mumu}
(\mu_1^2+\mu_2^2)=\alpha(\mu_1^2-\mu_2^2),
\end{equation}
with $\alpha=\frac{(\sqrt{\lambda_1}-\sqrt{\lambda_2})^2}{\lambda_1-\lambda_2}=\frac{(\sqrt{\lambda_1}-\sqrt{\lambda_2})}{\sqrt{\lambda_1}
+\sqrt{\lambda_2}}$, and $|\alpha|<1$. The former implies
\begin{equation}
\label{144}
\sqrt{\lambda_1}\mu_2^2+\sqrt{\lambda_2}\mu_1^2=0.
\end{equation}
Using (\ref{135}) and (\ref{136}) together with the condition (\ref{138}), we have two solutions. The first one is
\begin{equation}
K_0=\frac{\mu_1^2(\sqrt{\lambda_1}+\sqrt{\lambda_2})}{2(\lambda_1\sqrt{\lambda_2}-\lambda_2\sqrt{\lambda_1})}
=-K_3,
\end{equation}
where $K_0>0$ if $\mu_1^2>0,\; \lambda_1>\lambda_2$, or $\mu_1^2<0,\; \lambda_1<\lambda_2$.

The second solution is
\begin{equation}
K_0=\frac{\mu_1^2(\sqrt{\lambda_1}+\sqrt{\lambda_2})^2}{2(\lambda_1\lambda_2-\lambda_1^2)}=
K_3,
\end{equation}
where $K_0>0$ if $\mu_1^2>0,\; \lambda_2>\lambda_1$, or $\mu_1^2<0,\; \lambda_2<\lambda_1$.
 \end{description}

%%%%%%%%%%%%%%%%%%%%%%%%%%%%%%%%%%%%%%%%%%%%%%%%%%%%%%%%%%%%%%%%%%%%%%%%%%%%%%%%%%%%%%%%%

\section{\label{apendice3}EWSB in the case $w_0>0$}
It still remains to see if the economical 3-3-1 model is consistent, when the global minimum is found at
$K_0=|\boldsymbol K|$, i.e. if it is related to the Langrange multiplier $w_0>0$ (this situation was addressed in section~\ref{sc1}). In this case the vacuum expectation vectors $\langle\phi_1\rangle$ and $\langle\phi_2\rangle$ become linearly dependent, which implies that either $V_1=v_1=0$ or $v_2=0$ (cases where the electric charge generator is broken are not considered).

Following a similar approach to the one presented in Sect.~\ref{sec:sec6}, we analyze the second order term of the scalar potential, the one responsible to provide with masses to the physical Higgs fields. This term takes the form
\begin{equation}
V_{\{2\}}=\tilde{\boldsymbol K}^T_{\{1\}}\:\tilde E\: \tilde{\boldsymbol K}_{\{1\}}+2w_0\:\tilde{\boldsymbol K}^T_{\{0\}}\:\tilde g\:\tilde{\boldsymbol K}_{\{2\}}.
\end{equation}
Let us examine the two possible cases:
\begin{itemize}

\item $V_1=v_1=0:$ in this case all particles are decoupled. There are a total of six massive scalar particles with masses given by
\begin{equation}
\begin{split}
 M_{H_1}^2&=2w_0v_2^2,\:M_{H_2}^2=2\lambda_2v_2^2,\:M_{H_1^\prime}^2=2w_0v_2^2,\\
\:M_{A_1}^2&=2w_0v_2^2,\:M_{\phi_1^+(\phi_1^-)}^2=\lambda_4v_2^2/2,
\end{split}
\end{equation}
leaving the model with only six Goldstone bosons, which are not enough to provide with masses to the eight
gauge bosons associated to the same number of broken generators present in 3-3-1 models.
\item $v_2=0:$ for the notation established in~(\ref{136b}) we have
\begin{equation}
 \mathcal M^2_{neutral}=
\begin{pmatrix}
 2\lambda_1V_1^2& 0& 2\lambda_1v_1V_1\\
0&2w_0(v_1^2+V_1^2)&0\\
2\lambda_1v_1V_1&0&2\lambda_1v_1^2
\end{pmatrix},
\end{equation}
where $m_{H_2}^2=2w_0(v_1^2+V_1^2)$. The remaining submatrix has null determinant. In this way a total of two massive CP-even particles show up. For the CP-odd sector a massive particle $M_{A_2}^2=2w_0(v_1^2+V_1^2)$ is found.

In the charged sector we have
{\scriptsize

\begin{equation}
 \mathcal M^2_{charged}=
\begin{pmatrix}
0&0&0\\
0&2w_0(v_1^2+V_1^2)+\lambda_4v_1^2/4& \lambda_4v_1V_1/2\\
0&\lambda_4v_1V_1/2& 2w_0(v_1^2+V_1^2)+\lambda_4V_1^2/4
\end{pmatrix},
 \end{equation}}
where at least two additional massive charged particles are present, for a total of five massive particles; there remaining in this way seven Goldstone bosons, which is not enough to implement the Higgs mechanism in a consistent way.

 \end{itemize}

%%%%%%%%%%%%%%%%%%%%%%%%%%%%%%%%%%%%%%%%%%%%%%%%%%%%%%%%%%%%%%%%%%%%%%%%%%%%%%%%%%%%%%%%%

%%%%%%%%%%%%%%%%%%%%%%%%%%%%%%%%%%%%%%%%%%%%%%%%%%%%%%%%%%%%%%%%%%%%%%%%%%%%%%%%%%%%%%%%%

\end{document}